\documentclass[12pt]{article}
\usepackage{amsmath,amssymb,epsfig,bbold}

\newcommand{\beq}{\begin{equation}}
\newcommand{\be}{\begin{equation}}
\newcommand{\ee}{\end{equation}}
\begin{document}
\topmargin 0pt
\oddsidemargin=-0.4truecm
\evensidemargin=-0.4truecm
%\newpage
\setcounter{page}{1}
\begin{titlepage}
\vspace*{-2.0cm}   
%%%\vspace*{-1.0cm}
\begin{flushright}
hep-ph/0310119
\end{flushright}
\vspace*{0.5cm}
\begin{center}
\vspace*{0.2cm}  
%%%
{\Large \bf 
Supernova prompt neutronization neutrinos 
\vspace*{0.1cm}   
and neutrino magnetic moments}
\\
\vspace{1.0cm}
{\large Evgeny Kh. Akhmedov$^{\,(a)}$\footnote{On leave from the 
National
Research Centre Kurchatov Institute, Moscow, Russia. E-mail address:
akhmedov@ictp.trieste.it} and 
Takeshi Fukuyama$^{\,(b)}$\footnote{E-mail address: 
fukuyama@se.ritsumei.ac.jp}}
%\vspace{0.1cm}
\vspace{0.2cm}

{\em $^{(a)}$The Abdus Salam International Centre for Theoretical
Physics \\
Strada Costiera 11, I-34012 Trieste, Italy}

\vspace*{0.2cm}

{\em $^{(b)}$Department of Physics, Ritsumeikan University \\ 
Kusatsu, Shiga 525-8577, Japan}
\end{center}  
\vglue 1.5truecm

\begin{abstract}
It is shown that the combined action of spin-flavor conversions 
of supernova neutrinos due to the interactions of their Majorana-type 
transition magnetic moments with the supernova magnetic fields and 
flavor conversions due to the mass mixing can lead to the 
transformation of $\nu_e$ born in the neutronization process into 
their antiparticles $\bar{\nu}_e$. Such an effect would have a clear 
experimental signature and its observation would be a smoking gun evidence 
for the neutrino transition magnetic moments. It would also signify the 
leptonic mixing parameter $|U_{e3}|$ in excess of $10^{-2}$. 
\end{abstract}

\vspace{1.cm}
%%\centerline{Pacs numbers: .....}
\vspace{.3cm}
%\centerline{Keywords: sun, maf}
%\vspace{.3cm}

\end{titlepage}
\renewcommand{\thefootnote}{\arabic{footnote}}
\setcounter{footnote}{0}
\newpage

%%%%%%%%%%%%%%%%%%%%%%%%%%%%%%%%%%%%%%%%%%%%%%%%%%%%%%%%%%%%%%%%%%%%%%%%%%%%%
\section{Introduction}
%%%%%%%%%%%%%%%%%%%%%%%%%%%%%%%%%%%%%%%%%%%%%%%%%%%%%%%%%%%%%%%%%%%%%%%%%%%%%

%It is well known that 
Type-II supernovae explosions are accompanied by copious production 
of neutrinos and antineutrinos which carry away about 99\% of the 
emitted energy \cite{Suzuki}.
%is emitted in the form of neutrinos and antineutrinos 
%of all three species. 
The supernova (SN) neutrino flux consists of two main components: a 
very short ($\sim 10$ msec) pulse of $\nu_e$ produced in the process 
of neutronization of the SN matter which is followed by a longer ($\sim 10$ 
sec) pulse of thermally produced $\nu_e$, $\nu_\mu$, $\nu_\tau$ and their 
antiparticles (see fig. 1). Neutrino mixing, the convincing evidence for 
which was obtained in the solar, atmospheric and reactor neutrino 
experiments, results in the flavor conversions of SN neutrinos in 
supernovae and inside the Earth (for recent discussions see, e.g., 
\cite{DS}). In these transitions matter enhancement of neutrino oscillations 
(the MSW effect \cite{MSW}) plays an important role. 

Since the lepton flavor is not conserved, neutrinos should possess not only 
mass but also flavor-off-diagonal (transition) magnetic moments $\mu_{ab}$, 
which for Majorana neutrinos are the only allowed type of magnetic moments. In 
transverse magnetic fields, such magnetic moments would lead to a simultaneous 
rotation of neutrino spin and transformation of their flavor (spin-flavor 
precession) \cite{SchVal,VVO}. The neutrino spin-flavor precession can be 
resonantly enhanced by matter \cite{LM,Akh1,Akh2}, much in the same way as 
matter enhances neutrino oscillations. The neutrino resonance spin-flavor 
precession (RSFP) in the supernova magnetic fields can lead to transmutations 
of different supernova neutrino species \cite{AB,RSFP-SN,SNmodeldep}. 
For Majorana neutrinos, the possible conversions are 
\be
\nu_e\leftrightarrow \bar{\nu}_{\mu,\tau}\,,\quad 
\bar{\nu}_e\leftrightarrow \nu_{\mu,\tau}\,,\quad 
\nu_\mu\leftrightarrow \bar{\nu}_{\tau}\,,\quad 
\bar{\nu}_\mu\leftrightarrow \nu_{\tau}\,. 
\ee
At the same time, matter-enhanced neutrino flavor conversions 
\cite{MSW} lead to the transmutations 
\be
\nu_e\leftrightarrow \nu_{\mu,\tau}\,,\quad 
\bar{\nu}_e\leftrightarrow \bar{\nu}_{\mu,\tau}\,. 
\ee

%%%%%%%%%%%%%%%%%%%%%%%%%%%%%%%%%%% &&& %%%%%%%%%%%%%%%%%%%%%%%%%%%
\begin{figure}[tbh]
\begin{center}
\epsfig{file=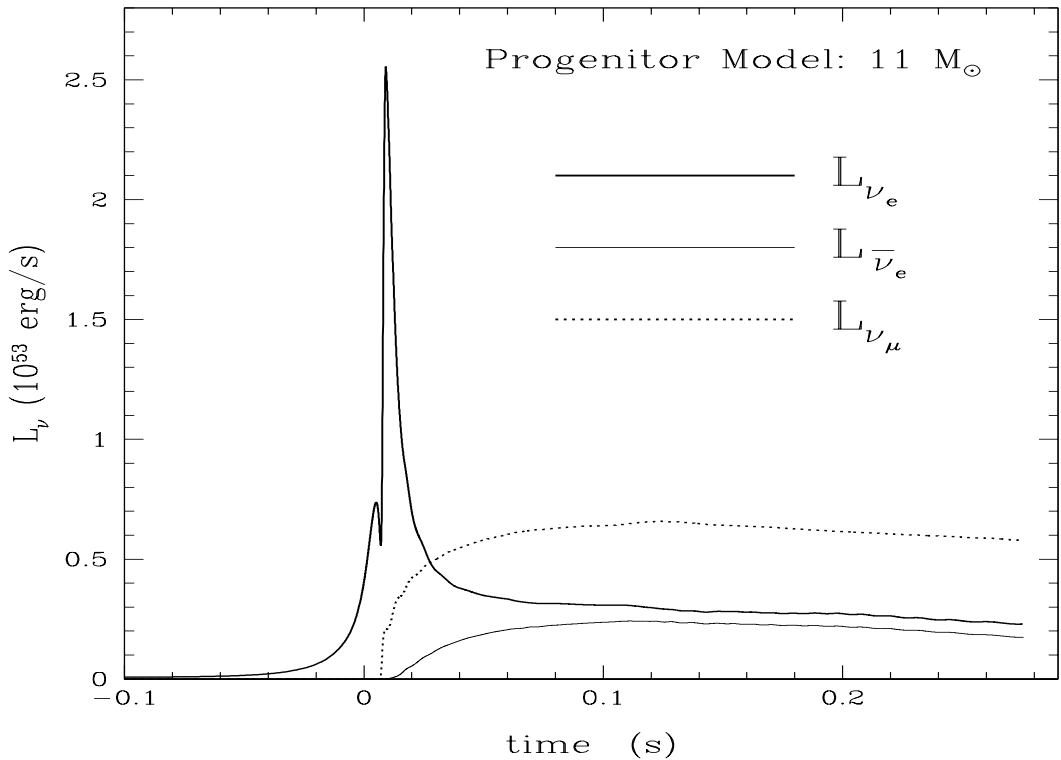,width=14cm,height=10cm}
\end{center}
\caption{\small Luminosities of the originally produced SN neutrinos 
as functions of time \cite{ThBuPi}. $L_{\nu_\mu}$ stands for the 
collective  luminosity of $\nu_\mu$, $\bar{\nu}_\mu$, $\nu_\tau$ and 
$\bar{\nu}_\tau$.}
\label{fig1}
\end{figure}
%%%%%%%%%%%%%%%%%%%%%%%%%%%%%%%%%%% &&& %%%%%%%%%%%%%%%%%%%%%%%%%%%

It is expected that the spectra of thermally produced SN neutrinos are 
characterized by the different mean energies \cite{Suzuki}: 
\be
\bar{E}_{\nu_e}\simeq 11~{\rm MeV}\,,\quad
\bar{E}_{\bar{\nu}_e}\simeq 16~{\rm MeV}\,,\quad
\bar{E}_{\nu_\mu}\simeq \bar{E}_{\bar{\nu}_\mu}\simeq 
\bar{E}_{\nu_\tau}\simeq \bar{E}_{\bar{\nu}_\tau}\simeq
25~{\rm MeV}\,.
\ee
Therefore the transitions between the neutrinos of electron and 
non-electron flavors, whether due to the MSW effects or due to the 
RSFP, should modify the spectra of the SN neutrinos observed at the 
Earth. 

Unfortunately, for thermally produced neutrinos, it is not 
possible to experimentally discriminate between the two effects. 
If, for example, electron neutrinos (or antineutrinos) are detected 
and their energy is found to be higher than expected, this can be 
due to the transition from either $\nu_{\mu,\tau}$ or  
$\bar{\nu}_{\mu, \tau}$; since these initial-state neutrinos have 
approximately the same energies, one cannot tell whether the 
observed hard spectrum results from the RSFP transitions of eq. (1) 
or the MSW transitions of eq. (2). Analogously, if the non-electronic 
flavor neutrinos or antineutrinos are detected and softer than 
expected spectrum is observed (e.g., the spectrum of the original 
$\nu_e$'s), this again can be due to either RSFP or MSW transitions.  
One cannot experimentally discriminate between the two possibilities 
because low-energy $\nu_x$ and $\bar{\nu}_x$ ($x=\mu, \tau$) can 
only be detected via neutral-current reactions which cannot tell low 
energy neutrinos from antineutrinos 
%%%%%%%%%%%%%%%%%%%%%%%%% &&& %%%%%%%%%%%%%%%%%%%%%%%%%%%%%%%%%%%%%%%%
%\footnote{Since no charged leptons are produced one cannot 
%distinguish between $\nu_x$ and $\bar{\nu}_x$ directly. The cross 
%sections of neutral current reactions with $\nu_x$ and $\bar{\nu}_x$ 
%are rather close to each other at low energies; therefore, given 
%the uncertainties of the SN neutrino fluxes and the possibility of 
%conversions between different neutrino species, one cannot discriminate 
%between the detected $\nu_x$ and $\bar{\nu}_x$ on the basis of the 
%observed event numbers.}.
%%%%%%%%%%%%%%%%%%%%%%%%% &&& %%%%%%%%%%%%%%%%%%%%%%%%%%%%%%%%%%%%%%%%

The situation with the prompt neutronization neutrinos is completely 
different. At the neutronization stage, the emitted neutrino flux 
consists almost entirely of $\nu_e$, the admixtures of other 
neutrino and antineutrino species being very small (see fig. 1). 
Resonance flavor or spin-flavor conversions, acting 
separately, can transform these neutrinos into, e.g., $\nu_\tau$ or 
$\bar{\nu}_\tau$ respectively; as was already pointed out, one cannot 
discriminate experimentally between these two possibilities. However, 
as we discuss below, the combined action of the MSW and RSFP 
transitions can result in the conversion $\nu_e\rightarrow \bar{\nu}_e$, 
leading to a detectable flux of prompt neutronization neutrinos in the 
$\bar{\nu}_e$ channel. Electron antineutrinos can be cleanly 
distinguished experimentally from all the other neutrino species and in 
fact are the easiest to detect. Thus, such an effect would have a 
very clear experimental signature: a short ($\sim 10$ msec) pulse of 
$\bar{\nu}_e$ preceding the longer pulse of thermal neutrinos of 
all species. The detection of such a signal would constitute an 
unambiguous evidence for neutrino magnetic moments.  

In the present paper we study the conversion of SN neutronization 
$\nu_e$ into $\bar{\nu}_e$ in the cases of the normal and inverted neutrino 
mass hierarchies in the full 3-flavor framework. In particular, we show 
that 3-flavor effects result in new spin-flavor resonances, absent in the 
2-flavor approximations. We consider these resonances in detail and study 
their role in the $\nu_e\to \bar{\nu}_e$ conversions in SN.

\section{Neutrino conversions in supernovae}
We assume neutrinos to be Majorana particles and consider the evolution 
of different neutrino species due to their flavor mixing and interaction of 
their transition magnetic moments with the SN magnetic fields. The 
neutrino 
evolution equation is 
\be
i\frac{d}{dr} \nu = H \nu
\label{evol1}
\ee
where $\nu=(\nu_e, \nu_\mu, \nu_\tau, \bar{\nu}_e,\bar{\nu}_\mu, 
\bar{\nu}_\tau)^T$ is the neutrino vector of state, $r$ is the coordinate  
along the neutrino trajectory  and $H$ is the effective Hamiltonian: 
\be
H=\left(
\begin{array}{cc}
M^2/2E   &  B(r) \\
 -B(r)  &  M^2/2E
\end{array}
\right) + V(r) \,.
\ee
In this equation $E$ is the neutrino energy, 
\be
%\frac{M^2}{2E}
M^2=U\left(\begin{array}{ccc}
0  & 0 & 0 \\
0  & \Delta m_{21}^2 & 0 \\
0  &  0 & \Delta m_{31}^2
\end{array}
\right) U^\dagger\,, \quad\quad 
B=\left(
\begin{array}{ccc}
0  & \mu_{e\mu} & \mu_{e\tau} \\
-\mu_{e\mu}  & 0 & \mu_{\mu\tau} \\
- \mu_{e\tau} & -\mu_{\mu\tau} & 0
\end{array}
\right) B_\perp(r)\,,
\label{Ham}
\ee
where $\Delta m_{ij}^2$ are the neutrino mass squared differences, $U$ 
is the leptonic mixing matrix 
and $B_\perp$ is the transverse magnetic field strength. The matrix of 
effective potentials 
\be
V={\rm diag}(V_e,\, V_\mu\,, V_\tau,\, V_{\bar{e}},\, V_{\bar{\mu}},\, 
V_{\bar{\tau}})
\label{V}
\ee
describes the coherent interactions of neutrinos of different flavor with 
matter. At tree level one has
\be
V_e =-V_{\bar{e}}=\sqrt{2}G_F (N_e-N_n/2)\,, \qquad 
%=\frac{G_F}{\sqrt{2}m_N}\rho (3Y_e-1)\,, \\
V_\mu = V_{\tau}=-V_{\bar{\mu}}=-V_{\bar{\tau}}=\sqrt{2}G_F(-N_n/2)\,,
%=\frac{G_F}{\sqrt{2}m_N}\rho (Y_e-1)\,, 
\label{Va}
\ee
where $G_F$ is the Fermi constant, $N_e$ and $N_n$ are the electron 
and neutron number densities, respectively. Although the tree-level 
potentials of $\nu_{\mu}$ and $\nu_{\tau}$ coincide, in one-loop order 
a difference between them arises due to the difference between the 
masses of the corresponding charged leptons \cite{BLM}:
\be
V_{\tau\mu}=V_\tau-V_\mu
\simeq \frac{3}{2\pi^2}G_F^2 m_\tau^2\left[(N_e+N_n)\ln\left(
\frac{M_W}{m_\tau}\right)-N_e-\frac{2}{3}N_n\right]\,.
\label{Vtaumu}
\ee
Here $M_W$ and $m_\tau$ are the $W$-boson and $\tau$-lepton masses, 
respectively; for antineutrinos one has $V_{\bar{\tau}\bar{\mu}}=- 
V_{\tau\mu}$. The quantity $V_{\tau\mu}$ is very small compared to the 
tree-level potentials $V_a$ ($a=e,\, \mu,\, \tau)$: 
$V_{\tau\mu}\sim 10^{-5} V_a$. However, it becomes important at very high 
densities $\rho\sim 10^{7} - 10^{8}$ g/cm$^3$ \cite{BLM,DS,AkhLuSm}.

The leptonic mixing matrix $U$ depends on three mixing angles\footnote{
In general, it also depends on the Dirac-type CP-violating 
phase $\delta_{\rm CP}$ and two Majorana-type phases. However, the 
Majorana phases do not affect neutrino oscillations and spin-flavor 
precession; the Dirac phase does not influence the SN neutrino 
signal \cite{AkhLuSm}.} which, along with the mass squared differences 
$\Delta m_{21}^2$ and $\Delta m_{31}^2$, can be obtained from the 
low-energy neutrino data. For the standard parameterization of the 
matrix $U$, the analyses of the solar, atmospheric and reactor neutrino 
experiments yield the following $1\sigma$ allowed ranges \cite{data}: 
\begin{eqnarray}
&& \Delta m_{21}^2\;\equiv\;\Delta m_{sol}^2\;=7.1^{+1.2}_{-0.6}\times 
10^{-5}~{\rm eV}^2 \,, \qquad \tan^2\theta_{21}=0.41^{+0.08}_{-0.07}\,
\nonumber \\
&& |\Delta m_{31}^2|\equiv \Delta m_{atm}^2=2.0^{+0.4}_{-0.3}\times 
10^{-3}~{\rm eV}^2\,, \qquad\sin^2 2\theta_{23}>0.92 \, \nonumber \\
&& |U_{e3}|=\sin\theta_{13}<0.16 \qquad\qquad\qquad  
\label{exdata}
\end{eqnarray}
The data allow two possible hierarchies (orderings) of neutrino masses, 
the normal hierarchy ($\Delta m_{31}^2>0$) and inverted hierarchy 
($\Delta m_{31}^2<0$). In the case of the normal hierarchy 
the third mass eigenstate $\nu_3$, which is separated by a larger mass 
gap from the other two and has a small admixture of $\nu_e$ ($|U_{e3}|^2\ll 
1$), is heavier than $\nu_1$ and $\nu_2$, whereas in the case of the 
inverted hierarchy $\nu_3$ is the lightest mass eigenstate. The SN 
neutrino transmutations depend crucially on the neutrino mass hierarchy.

The density dependence of the energy levels of neutrino matter eigenstates 
can be read off from the effective Hamiltonian $H$ in eq. (\ref{Ham}) 
\cite{DS,AkhLuSm}. It is given in figs. 2 and 3. At very high densities, 
when the potentials $V_a$ and the potential difference $V_{\tau\mu}$ are 
very large, the diagonal terms in $H$ dominate and matter eigenstates 
essentially coincide with the flavor eigenstates $\nu_e$, $\nu_\mu$ and 
$\nu_\tau$. In the intermediate density range $10^4~{\rm g/cm}^3\ll \rho 
\ll \rho_{\tau\mu}\sim 10^7 - 10^8~{\rm g/cm}^3$ one can neglect the 
potential difference $V_{\tau\mu}$ whereas $V_a$ are still very large. In 
that density domain the matter eigenstates are $\nu_e$, $\nu_\mu'$ and 
$\nu_\tau'$, where $\nu_\mu'$ and $\nu_\tau'$ are the states that 
diagonalize the $\mu-\tau$ sector of the effective Hamiltonian $H$ in 
vacuum. In the density region $\rho\lesssim 10^6~{\rm g/cm}^3$ a number of 
flavor and spin-flavor conversions occur (see figs. 2 and 3). As the 
matter density further decreases and approaches zero, the matter eigenstates 
go into the mass eigenstates $\nu_1$, $\nu_2$ and $\nu_3$. 

%%%%%%%%%%%%%%%%%%%%%%%%%%%%%%%%%%% &&& %%%%%%%%%%%%%%%%%%%%%%%%%%%
\begin{figure}[tbh]
\begin{center}
\epsfig{file=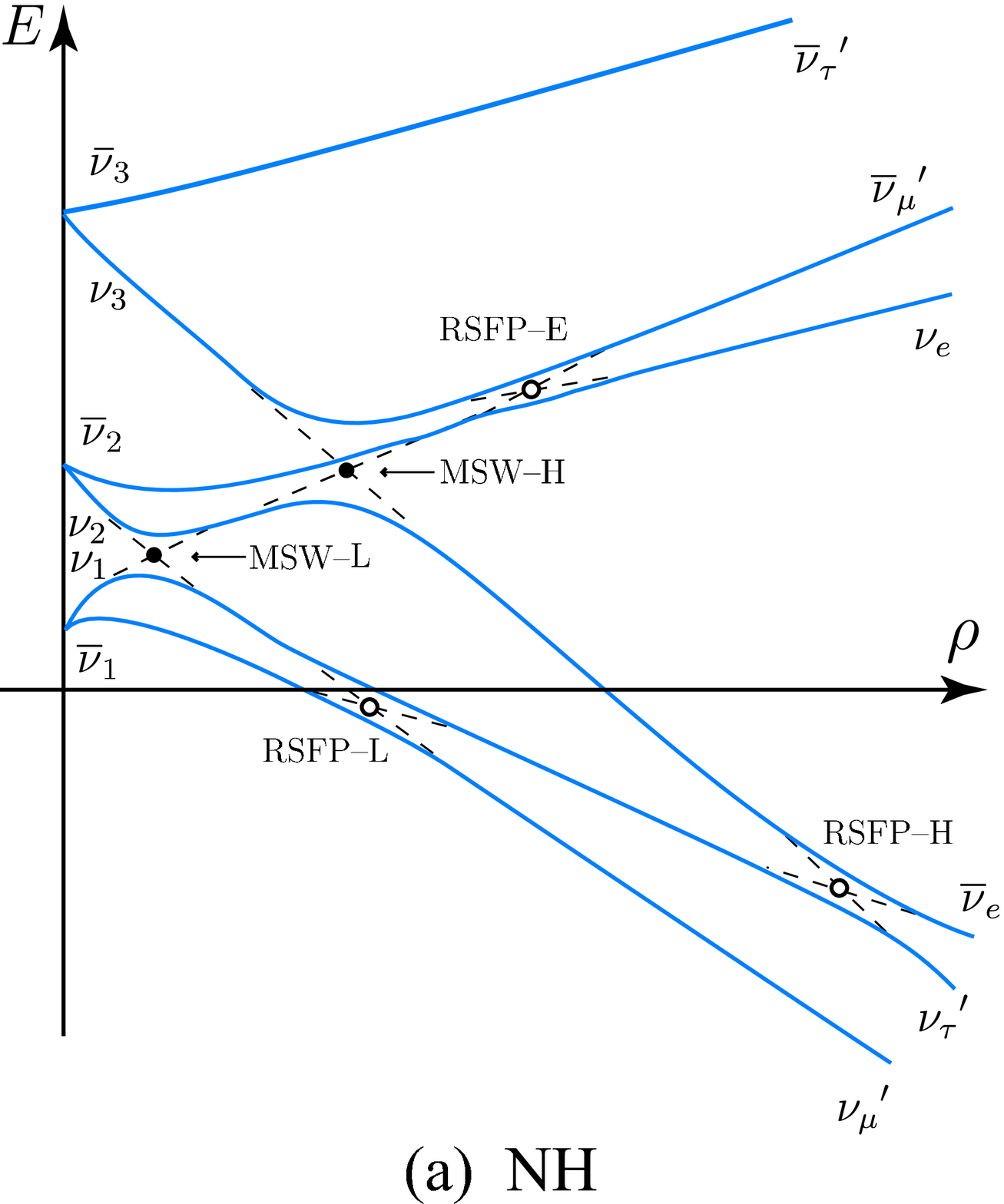,width=7.8cm,height=9.0cm}
\hspace*{0.4cm}
\epsfig{file=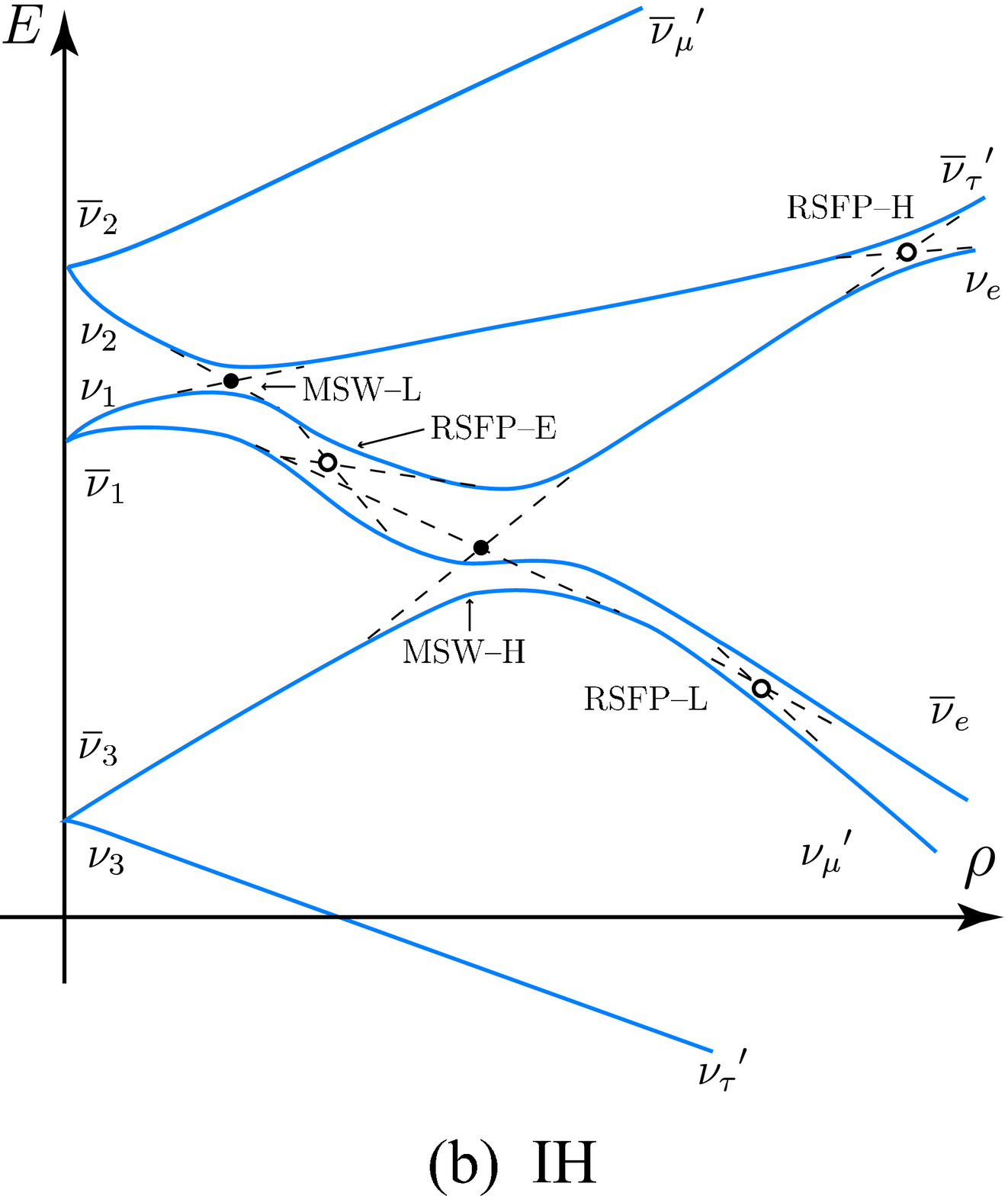,width=7.8cm,height=9.0cm}
\end{center}
\vspace*{-5mm}
\caption{\small Neutrino energy level schemes in SN for the normal (a) and 
inverted (b) mass hierarchies. The case $s_{13}>(s_{13})_c$. }
\label{fig2}
\end{figure}
%%%%%%%%%%%%%%%%%%%%%%%%%%%%%%%%%%% &&& %%%%%%%%%%%%%%%%%%%%%%%%%%%

The resonances occur when the separation between the 
energy levels of neutrino matter eigenstates become minimal. In the 
two-flavor approximation the resonances correspond to the points where the 
diagonal elements of the effective Hamiltonian $H$ become pairwise equal 
(level crossing points). 
We summarize now briefly the resonance transitions and the 
corresponding resonance conditions  
(the suffixes ``H'' and ``L'' stand for the high and low resonance 
densities corresponding to $\Delta m_{31}^2$ and $\Delta m_{21}^2$, 
respectively): 
 
$\bullet$ RSFP-H: ~~~$\bar{\nu}_{e}\leftrightarrow \nu_\tau'$ (normal 
hierarchy);~~~$\nu_{e}\leftrightarrow \bar{\nu}_\tau'$ (inverted 
hierarchy);  
\be
\sqrt{2}\,G_F\frac{1}{m_N}\,\rho_{res} (1-2Y_e)\simeq 
\frac{|\Delta m_{31}^2|}{2E}\cos 2\theta_{13}\,;
\label{res1}
\ee

$\bullet$ RSFP-L: ~~~$\bar{\nu}_{e}\leftrightarrow \nu_\mu'$ 
(normal and inverted hierarchies);
\be
\sqrt{2}\,G_F\frac{1}{m_N}\,\rho_{res} (1-2Y_e)\simeq 
\frac{\Delta m_{21}^2}{2E}\cos 2\theta_{12}\,;
\label{res2}
\ee

$\bullet$ RSFP-X:~~~$\bar{\nu}_\mu'\leftrightarrow \nu_\tau'$ (normal 
hierarchy);~~~$\nu_\mu'\leftrightarrow \bar{\nu}_\tau'$ (inverted 
hierarchy);  
\be
\sqrt{2}\,G_F\frac{1}{m_N}\,\rho_{res} (1-Y_e)\simeq 
\frac{|\Delta m_{31}^2|}{2E}\cos^2\theta_{13}\,;
\label{res3}
\ee
These resonances occur when $\sin\theta_{13}$ satisfies 
\footnote{We use the standard notation $s_{ij}\equiv\sin\theta_{ij}$, 
$c_{ij}\equiv \cos\theta_{ij}$.}
\be
s_{13} < \cos 2\theta_{12}\,
\frac{\Delta m_{21}^2}{|\Delta m_{31}^2|}\simeq 0.015\,.
\label{cond1}
\ee

$\bullet$ RSFP-E: ~~~$\nu_{e}\leftrightarrow \bar{\nu}_\mu'$ (normal 
hierarchy);~~~$\bar{\nu}_{e}\leftrightarrow \nu_\mu'$ (inverted hierarchy);  
\be
\sqrt{2}\, G_F \frac{1}{m_N}\,\rho_{res} \simeq 
\left[(1 \mp a+z) \pm \sqrt{(1 \mp a+z)^2 \pm 4a}\right]
\frac{|\Delta m_{31}^2|}{2E}\,.
\label{res6}
\ee
Here the upper and lower signs refer to the normal and  inverted mass 
hierarchies, respectively, and 
\be
a=\frac{Y_e}{1-2Y_e}\,\frac{\Delta m_{21}^2}{|\Delta m_{31}^2|}\,
\cos 2\theta_{12}\,,\quad\quad
z=s_{13}\,\frac{Y_e}{1-2Y_e}\,.
\label{not1}
\ee 
The resonances occur when 
\be
s_{13} > (s_{13})_c\equiv \cos 2\theta_{12}\,
\frac{\Delta m_{21}^2}{|\Delta m_{31}^2|}\simeq 0.015\,,
\label{cond2}
\ee
which is the opposite condition to (\ref{cond1}). Note that 
$(s_{13})_c=s_{13}(a/z)$. 

$\bullet$ MSW-H: ~~~$\nu_{e}\leftrightarrow \nu_\tau'$ (normal hierarchy);
~~~$\bar{\nu}_{e}\leftrightarrow \bar{\nu}_\tau'$ (inverted hierarchy);  
\be
\sqrt{2}\,G_F\frac{1}{m_N}\,\rho_{res} Y_e\simeq 
\frac{|\Delta m_{31}^2|}{2E}\cos 2\theta_{13}\,;
\label{res4}
\ee

$\bullet$ MSW-L: ~~~$\nu_{e}\leftrightarrow \nu_\mu'$ 
(normal and inverted hierarchies);
\be
\sqrt{2}\,G_F\frac{1}{m_N}\,\rho_{res} Y_e\simeq 
\frac{\Delta m_{21}^2}{2E}\cos 2\theta_{12}\,.
\label{res5}
\ee
In eqs. (\ref{res1}) -- (\ref{res5}) $m_N$ is the nucleon mass and 
$Y_e$ is the number of electrons per nucleon, $Y_e=N_e/(N_e+N_n)$. 
Note that the above resonances occur in the so-called isotopically 
neutral region of SN, where $Y_e$ is slightly below 0.5 and very close 
to this value: $1-2Y_e\sim 10^{-4} - 10^{-3}$. For this reason the 
RSFP-H and RSFP-L resonance densities are about three orders of magnitude 
higher than those of the corresponding MSW resonances. Because of the 
smallness of $1-2Y_e$, the slopes of the energy levels of $\nu_e$, 
$\bar{\nu}_\mu'$ and $\bar{\nu}_\tau'$ (and also those of $\bar{\nu}_e$, 
$\nu_\mu'$ and $\nu_\tau'$ ) as functions of matter density are almost 
identical, see figs. 2 and 3; this, in turn, leads to a number of subtle 
effects which we discuss below.  

%%%%%%%%%%%%%%%%%%%%%%%%%%%%%%%%%%% &&& %%%%%%%%%%%%%%%%%%%%%%%%%%%
\begin{figure}[tbh]
\begin{center}
\epsfig{file=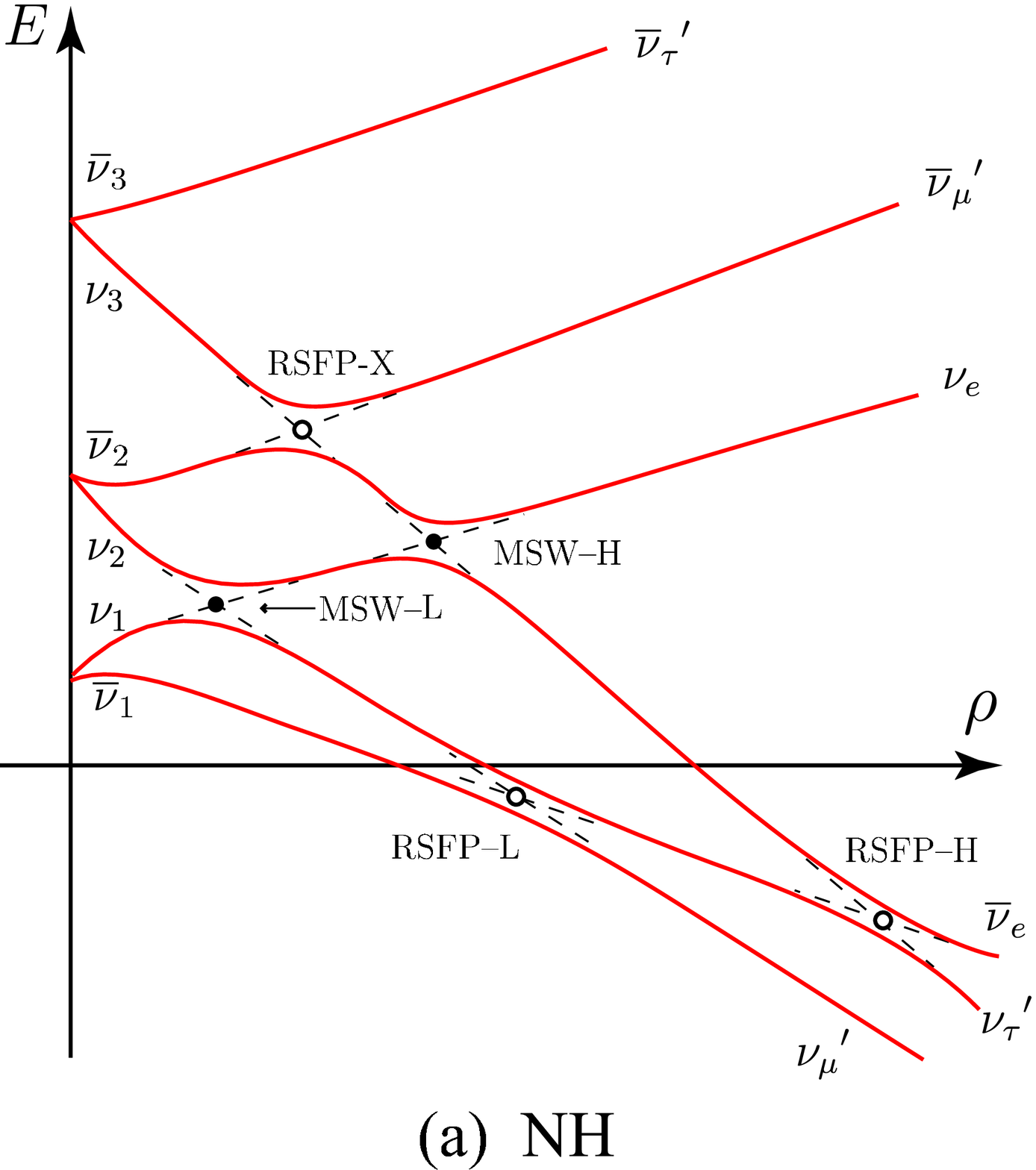,width=7.8cm,height=9.0cm}
\hspace*{0.4cm}
\epsfig{file=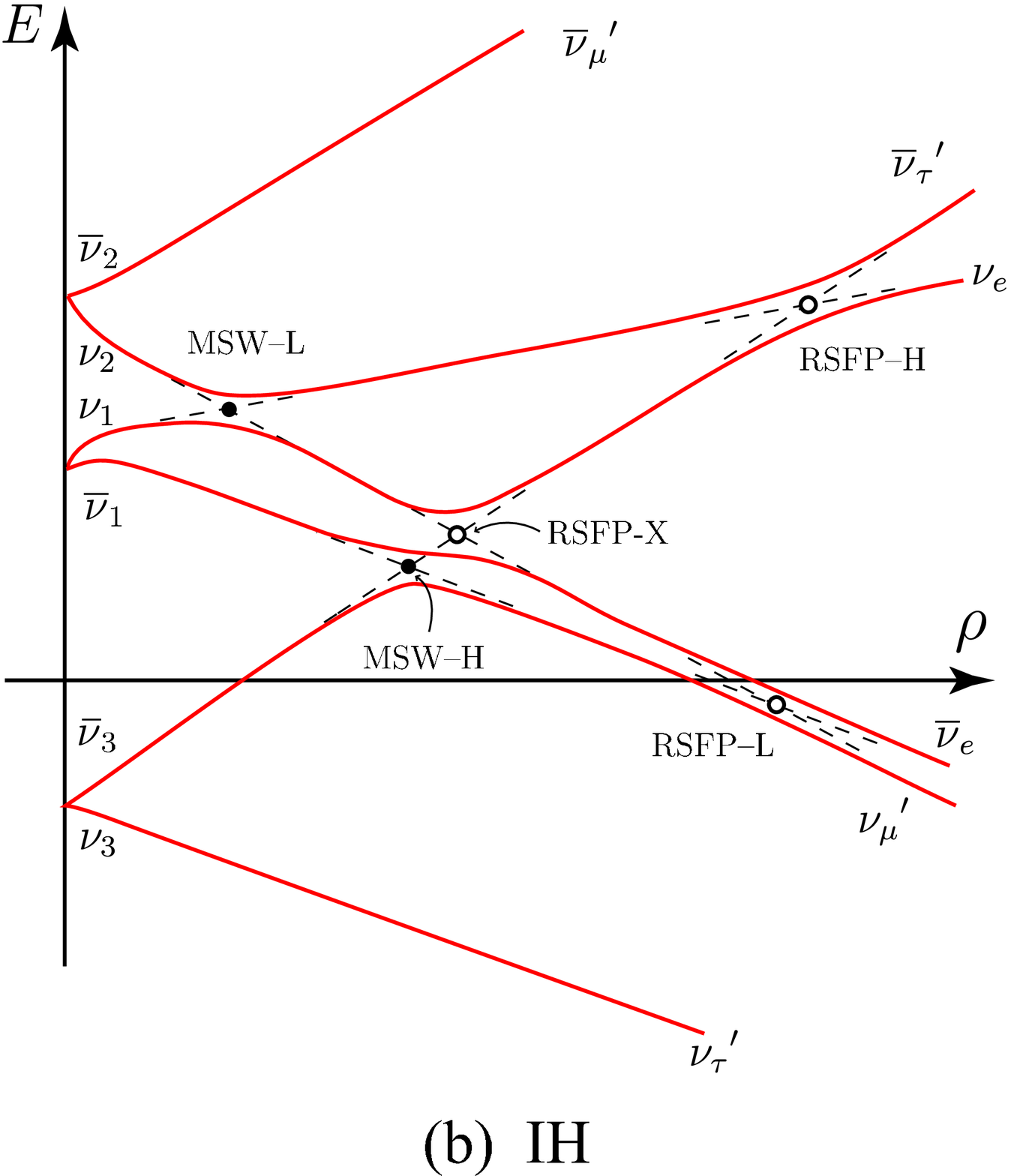,width=7.8cm,height=9.0cm}
\end{center}
\vspace*{-5mm}
\caption{\small Neutrino energy level schemes in SN for the normal (a) and 
inverted (b) mass hierarchies. The case $s_{13}<(s_{13})_c$. }
\label{fig3}
\end{figure}
%%%%%%%%%%%%%%%%%%%%%%%%%%%%%%%%%%% &&& %%%%%%%%%%%%%%%%%%%%%%%%%%%

The spin-flavor precession resonances RSFP-H and RSFP-L have been widely 
discussed in the literature \cite{AB,RSFP-SN}; the RSFP-X resonance was 
considered in ref. \cite{AB}. The \mbox{RSFP-E} resonances of eq. (\ref{res6}) 
\footnote{ The suffix ``E'' indicates that electron-type neutrinos are 
involved.} 
are new and considered here for the first time. They are a pure 
\mbox{3-flavor} effect. In the 2-flavor approach, the RSFP transitions in SN 
between the electron-type neutrinos or antineutrinos and 
$\bar{\nu}_\mu'$($\nu_\mu'$) are expected to occur only in the $\bar{\nu}_e 
\leftrightarrow \nu_\mu'$ channel and to be driven by the ``solar'' mass 
squared difference $\Delta m_{21}^2$ and mixing angle $\theta_{12}$ [see eq. 
(\ref{res2})]. The reason for this is that the $\nu_e$ components in the mass 
eigenstates $\nu_1$ and $\nu_2$ are large, whereas the weight of $\nu_e$ in 
the mass eigenstate $\nu_3$ is small (equal to $s_{13}^2$). However, even 
this small weight can cause, in the case of the normal mass hierarchy, 
resonance $\nu_e\leftrightarrow \bar{\nu}_\mu'$ transitions at certain 
densities (see fig. 2a).  
If the neutrino mass hierarchy is inverted, the RSFP-E resonance takes place 
in the $\bar{\nu}_e\leftrightarrow \nu_\mu'$ channel, but at a density which 
is different from the RSFP-L one and is given by eq. (\ref{res6}) (fig. 2b). 
Thus, in that case the $\bar{\nu}_e\leftrightarrow \nu_\mu'$ conversions 
can occur in  two different density regions. 

We now summarize briefly the main features of the RSFP-E resonances; 
more detailed discussion of their properties will be given in sec. 3.
The RSFP-E resonance densities (\ref{res6}) depend crucially on the ratio 
of the two small parameters, $s_{13}^2$ and $1-2Y_e$. They also depend on 
both the ``atmospheric'' and ``solar'' mass squared differences 
$\Delta m_{31}^2$ and $\Delta m_{21}^2$. The dependence on the small 
$\Delta m_{21}^2$ is important since it is enhanced by the small $1-2Y_e$ in 
the denominator of the parameter $a$ in (\ref{not1}).

The RSFP-E resonances take place only when $s_{13}$ exceeds the critical 
value $(s_{13})_c$ defined in (\ref{cond2}). In the case of the normal mass 
hierarchy, the RSFP-E resonance density $(\rho_{res})_{\rm RSFP-E}$ is larger 
than the MSW-H one, $(\rho_{res})_{\rm MSW-H}$. With decreasing $s_{13}$,  
$(\rho_{res})_{\rm RSFP-E}$ decreases and approaches 
$(\rho_{res})_{\rm MSW-H}$; the two densities coincide when 
$s_{13}=(s_{13})_c$. 
\footnote{Note that the RSFP-E resonance condition (\ref{res6}) is 
approximate; though its accuracy is very good (better than 3\%), it is  
still not sufficient to find the critical value $(s_{13})_c$. The latter 
is obtained from a more precise resonance condition, see discussion in 
sec. 3.} 
For smaller $s_{13}$, in the limit $B_\perp \to 0$  
the $\bar{\nu}_\mu'$ level crosses 
the matter eigenstate level $\nu_{1m}$ (which coincides with the $\nu_e$   
level at very high densities) below the MSW-H resonance point, where 
$\nu_{1m}$ corresponds to $\nu_\tau'$ rather than to $\nu_e$. 
Thus, when $s_{13}$ becomes smaller than $(s_{13})_c$, the \mbox{RSFP-E} 
resonance transforms into the RSFP-X one, see fig. 3a.
Similar situation takes place in the case of the inverted neutrino mass 
hierarchy, except that the RSFP-E transition is in the $\bar{\nu}_e
\leftrightarrow \nu_\mu'$ channel, and with $s_{13}$ decreasing and 
approaching $(s_{13})_c$ from above the RSFP-E resonance density 
increases and approaches $(\rho_{res})_{\rm MSW-H}$ from below. 
The inverted hierarchy case with $s_{13}<(s_{13})_c$ is illustrated by 
fig. 3b. It should be noted that the transformation of the \mbox{RSFP-E} 
resonances into the RSFP-X ones with decreasing $s_{13}$ is not sharp -- 
due to the finite width of the resonances there is a border region where 
these resonances coexist. We discuss this in more detail in sec. 4.

The resonance conditions discussed above, except those for the RSFP-E 
resonances, were found in the two-flavor approximation. This approximation is 
quite accurate for the \mbox{RSFP-H}, MSW-H and MSW-L resonances. For the 
RSFP-L transitions it is sufficiently accurate when $s_{13}$ is not too 
large: for $s_{13}< 0.05$ the accuracy is better than 15\%. At the same time,  
for $s_{13}\simeq 0.16$ which is the maximum allowed by the current data 
value, the 2-flavor approximation only gives the RSFP-L resonance density  
by an order of magnitude, and full 3-flavor consideration is necessary when 
more precise results are needed. The sensitivity of $(\rho_{res})_{\rm 
RSFP-L}$ to 3-flavor effects and in particular to the mixing parameter 
$s_{13}$ is to a large extent a consequence of the above-mentioned smallness 
of $1-2Y_e$. It is because of this smallness that the closely situated 
$\bar{\nu}_e$ and $\nu_\mu'$ energy levels have nearly the same slope and a 
small variation of $s_{13}$ or $\Delta m_{31}^2$ can lead to a sizeable 
change in the position of their crossing point, i.e. in 
$(\rho_{res})_{\rm RSFP-L}$. The accuracy of the 2-flavor approximation is 
in general slightly better in the case of the inverted mass hierarchy 
than in the normal hierarchy case. For the RSFP-E transitions, 3-flavor 
effects are crucial.

In addition to the above conversions, there are MSW transitions  governed 
by the effective potential difference $V_{\tau\mu}$ and ``atmospheric'' 
mass squared difference $\Delta m_{31}^2$. The corresponding level 
crossings occur in the antineutrino channel for the normal mass hierarchy 
and in the neutrino channel for the inverted hierarchy. The resonance 
densities $\rho_{res}$ lie in the interval $(0,\;\rho_{\tau\mu})$, their 
values depending on the value of $\theta_{23}$. The 
exact location of these resonances is, however, unimportant: the main 
role of the potential $V_{\tau\mu}$ is to suppress the flavor mixing at 
high densities and convert the neutrino states $\nu_\mu$ and $\nu_\tau$ 
into $\nu_\mu'$ and $\nu_\tau'$ at intermediate densities. The 
corresponding transitions (not shown in figs.~2 and 3) are 
\cite{DS,AkhLuSm}: 
\begin{eqnarray}
&& \nu_\mu\to \nu_\mu'\,, \quad \nu_\tau\to \nu_\tau'\,,\quad 
\bar{\nu}_\mu\to \bar{\nu}_\tau'\,, \quad \bar{\nu}_\tau\to 
\bar{\nu}_\mu'\,\qquad \mbox{(normal mass hierarchy)}\,; \\
&& \nu_\mu\to \nu_\tau'\,, \quad \nu_\tau\to \nu_\mu'\,,\quad  
\bar{\nu}_\mu\to \bar{\nu}_\mu'\,, \quad \bar{\nu}_\tau\to 
\bar{\nu}_\tau'\,\qquad \mbox{(inverted mass hierarchy)}\,. 
\end{eqnarray}

The evolution of neutrinos inside the SN can be readily followed in the 
adiabatic approximation (i.e. when the matter density changes slowly 
enough along the neutrino path). Since the flavor mixing is strongly 
suppressed at very high densities, neutrinos born as the flavor eigenstates 
$\nu_e$, $\nu_\mu$ and $\nu_\tau$ in the supernova's core are initially 
in matter eigenstates. In the adiabatic regime the transitions between 
different matter eigenstates are strongly suppressed and so the neutrino 
states just follow the evolution of matter eigenstates (solid lines in 
figs. 2 and 3) through a number of resonance conversions until they become 
mass eigenstates. If adiabaticity is strongly broken in some of the 
resonances, the neutrino state ``hops'' from the initial matter eigenstate 
onto the nearby one (in figs. 2 and 3, it follows the dashed line through 
the level crossing). In general, when the adiabaticity in the $i$th 
resonance is neither perfect nor badly broken, there are finite 
probabilities for the neutrino state to follow the initial matter 
eigenstates or to hop to another one ($1-P_i$ and $P_i$, respectively). 

In each of the MSW and RSFP resonances, the hopping probability $P_i$ is 
approximately given by
\be
P_i\simeq e^{-\frac{\pi}{2}\gamma_i}\,,
\label{hop}
\ee
where $\gamma_i$ is the corresponding adiabaticity parameter
\footnote{Eq. (\ref{hop}) has to be modified when the relevant vacuum 
mixing angle is relatively large. This, however, is unimportant for our 
discussion.}. 
For the MSW transitions one has 
\be
\gamma_{\rm MSW}=\frac{\sin^2 2\theta_{ij}}{\cos 2\theta_{ij}}
\frac{\Delta m_{ij}^2}{2E} L_\rho\,,
\label{gammaMSW1}
\ee
whereas for the RSFP-H, RSFP-L and RSFP-X transitions,  
\be
\gamma_{\rm RSFP}\simeq \frac{8E}{\Delta m_{ij}^2}(\mu_{ab}B_{\perp r})^2 
L_\rho\,.
\label{gammaRSFP1}
\ee
Here $B_{\perp r}$ is the transverse magnetic field strength at the relevant 
RSFP resonance and $L_\rho$ is the scale height of the effective matter 
density at the resonance: 
\be
L_\rho=\left|\frac{1}{\rho Y_{\rm eff}}\frac{d (\rho Y_{\rm eff})}{dr}
\right|_{res}^{-1}\,,
\ee
where $Y_{\rm eff}=Y_e$, $(1-2Y_e)$ or $(1-Y_e)$, depending on the 
resonance [see eqs. (\ref{res1}) -- (\ref{res5})]. 

The adiabaticity parameter of the RSFP-E conversion cannot in general be 
written in a simple form. Here we give it for the normal neutrino 
mass hierarchy in the limiting case $s_{13}\gg (s_{13})_c$ which will be 
useful for our later discussion:
\be
\gamma_{\rm RSFP-E}\simeq \frac{8 E \left(\mu_{e\mu'}\, B_{\perp r}\right)^2}
{s_{13}^2\,\Delta m_{31}^2-\cos 2\theta_{12}\,\Delta m_{21}^2}\,L_\rho\,.
\label{gammaRSFPE}
\ee
Notice that this parameter is enhanced with respect to the typical RSFP-X  
adiabaticity parameter (with $\mu_{\mu\tau}$ replaced by $\mu_{e\mu'}$) 
by the factor 
\be
[s_{13}^2-\cos 2\theta_{12} 
(\Delta m_{21}^2/\Delta m_{31}^2)]^{-1}\gtrsim 10^2\,.
\label{enh} 
\ee
In the adiabatic regime the adiabaticity parameters satisfy $\gamma_i\gg 1$ 
and the corresponding hopping probabilities $P_i$ are strongly suppressed. 

The adiabaticity parameters 
depend crucially on the SN matter density and magnetic field profiles. 
Numerical simulations yield the SN density profiles which for 
$r> 10~{\rm km}$ can be well approximated by 
\be
\rho(r)=\rho_0\left(\frac{10~{\rm km}}{r}\right)^3\,,
\label{rho}
\ee
with $\rho_0\simeq$ (1 -- 15)$\times 10^{13}$ g/cm$^3$. 
The SN magnetic field 
profiles are essentially unknown; in most of the studies of the SN 
implications of neutrino magnetic moments the profiles 
\be
B_\perp(r)=B_0\left(\frac{r_0}{r}\right)^k\,,
\label{B}
\ee
with $k=2$ or 3 were assumed \cite{AB,RSFP-SN}. The exponent $k=3$ 
corresponds to the dipole magnetic field, while $k=2$, to the magnetic flux 
conservation (frozen field). 
We use the profile (\ref{B}) with $r_0=10$ km, $k=2$ or 3 and $B_0$ a 
free parameter. 

It is convenient to express, using eqs. (\ref{exdata}) -- (\ref{res5}), 
the adiabaticity parameters (\ref{gammaMSW1}) and 
(\ref{gammaRSFP1}) in terms of the resonance densities: 
%Here we have used $L_\rho\simeq r_{res}/3$. 
%\begin{eqnarray}
\be
 \gamma_{\rm MSW}\simeq 55 \tan^2 2\theta_{ij}\, Y_e  
\left[\frac{\rho_0}{8\cdot 10^{13}~{\rm g/cm}^3}\right]^{1/3}\,
\left[\frac{\rho_{res}}{1\,{\rm g/cm}^3}\right]^{2/3}\,,
\label{gammaMSW2} \\
\ee
\be
 \gamma_{\rm RSFP}\simeq 
3.8\times 10^{-5}\frac{1}{Y_{\rm eff}}
\left[\frac{8\cdot 10^{13}~{\rm g/cm}^3}{\rho_0}\right]
\left(\frac{\rho_{res}}
{\rho_0}\right)^{\frac{2k-4}{3}}
\left[\frac{\mu_{ab}}{10^{-13}\mu_B}\,\frac{B_0}{10^{14}~{\rm G}}
\right] ^2\,.
\label{gammaRSFP2}
\ee
%\end{eqnarray}
Note that in the case $k=2$ the RSFP adiabaticity parameter does not 
depend on the resonance density, except possibly through the parameter 
$Y_{\rm eff}$. 

The resonance densities of various RSFP and MSW transitions can be readily 
obtained from eqs. (\ref{exdata}) -- (\ref{res5}).
For the typical SN neutrino energies $E\simeq 10$ -- 50 MeV one finds 
\be
(\rho_{res})_{\rm RSFP-H}\simeq (0.5 - 2.5)\times 10^6 ~{\rm 
g/cm}^3\,, \quad (\rho_{res})_{\rm RSFP-L}\simeq (0.8 - 4)\times 10^4  
~{\rm g/cm}^3\,, 
\label{resdens1}
\ee
\be
(\rho_{res})_{\rm MSW-H}\simeq (\rho_{res})_{\rm RSFP-X}\simeq 500 - 
2500 ~{\rm g/cm}^3\,, \quad (\rho_{res})_{\rm MSW-L}\simeq 8 - 40 
~{\rm g/cm}^3\,. 
\label{resdens2}
\ee
For neutrino energy $E=15$ MeV (average energy of the neutronization $\nu_e$) 
and normal neutrino mass hierarchy the RSFP-E resonance density varies between 
$1.7\times 10^3$ and $1.3\times 10^4$ g/cm$^3$, whereas for the inverted mass 
hierarchy it changes in the interval ($7\times 10^2$ -- $2\times 10^3$) 
g/cm$^3$, depending in the value of $s_{13}$. 

{}From eqs. (\ref{exdata}) and  (\ref{gammaMSW2}) we see that the MSW-L 
transitions are always adiabatic, while the MSW-H ones are adiabatic for 
$s_{13}^2 > 10^{-4}$ and non-adiabatic for $s_{13}^2< 10^{-5}$. From eq. 
(\ref{gammaRSFP2}) it follows that for the RSFP transitions to be adiabatic 
one needs
\begin{eqnarray}
\mbox{For}~k=2:\quad &&
\left[\frac{\mu_{ab}}{10^{-13}\mu_B}\,\frac{B_0}{10^{14}~{\rm G}}\right]
>5~~(\mbox{RSFP-H and RSFP-L})\,; 
\label{adiab1} \\
&& \left[\frac{\mu_{ab}}{10^{-13}\mu_B}\,\frac{B_0}{10^{14}~{\rm G}}\right]
>10^2~~(\mbox{RSFP-X})\,.
\label{adiab2} 
\end{eqnarray}
\begin{eqnarray}
\mbox{For}~k=3: &&
\left[\frac{\mu_{ab}}{10^{-13}\mu_B}\,\frac{B_0}{10^{14}~{\rm G}}\right]
>2\times 10^3~~(\mbox{RSFP-H})\,; 
\label{adiab3} \\ 
&& \left[\frac{\mu_{ab}}{10^{-13}\mu_B}\,\frac{B_0}{10^{14}~{\rm G}}\right]
>8\times 10^3~~(\mbox{RSFP-L})\,; 
\label{adiab4} \\
&& \left
[\frac{\mu_{ab}}{10^{-13}\mu_B}\,\frac{B_0}{10^{14}~{\rm G}}\right]
>4\times 10^5~~(\mbox{RSFP-X})\,.
\label{adiab5} 
\end{eqnarray}
The adiabaticity conditions for the RSFP-E resonances depend on the 
resonance density, which in turn is a sensitive function of $s_{13}$. For the 
case of interest to our study, normal mass hierarchy and $s_{13}$ ranging 
from $(s_{13})_c\simeq 0.015$ to $(s_{13})_{max}=0.16$, the adiabaticity 
conditions for the RSFP-E resonance are
\begin{eqnarray}
&& \left[\frac{\mu_{ab}}{10^{-13}\mu_B}\,\frac{B_0}{10^{14}~{\rm 
G}}\right] >  4 - 5\,, \qquad\qquad\qquad ~k=2\,; \label{condit1} \\
&& \left[\frac{\mu_{ab}}{10^{-13}\mu_B}\,\frac{B_0}{10^{14}~{\rm 
G}}\right] >  (0.7 - 1)\times 10^4\,, \quad\quad k=3\,,
\label{condit2}
\end{eqnarray}
where we have taken into account the enhancement factor (\ref{enh}) and 
the fact that $(\rho_{res})_{\rm RSFP-E}$ can exceed $(\rho_{res})_{\rm 
MSW-H}$ by up to a factor of 8. Conditions (\ref{condit1}) and 
(\ref{condit2}) can be relaxed if there is an accidental partial 
cancellation between the two terms in square brackets in (\ref{enh}).

Note that for the RSFP-H transitions $\mu_{ab}=\mu_{e\tau'}$, for 
the RSFP-L and RSFP-E transitions $\mu_{ab}=\mu_{e\mu'}$, and for the 
RSFP-X ones $\mu_{ab}=\mu_{\mu\tau}$. 

\section{RSFP-E conversions and overlap of resonances}

The RSFP-E resonance in general takes place at densities that are not very 
far from that of the MSW-H resonance, and the RSFP-X resonance is always very 
close to the MSW-H one. For this reason in certain ranges of $s_{13}$ some 
of these resonances may overlap. To study this phenomenon and also to get more 
insight into the nature of the RSFP-E resonance, the following approach proves 
to be useful. Consider the normal mass hierarchy case (the inverted hierarchy 
case is analyzed quite analogously). We note that in the vicinity  
of the MSW-H resonance the energy levels of $\nu_e$, $\nu_\tau'$ and 
$\bar{\nu}_\mu'$ are close to each other, whereas the other three energy 
levels are rather far. In this region one can therefore neglect the 
influence of the far-lying states and consider the evolution of $\nu_e$, 
$\nu_\mu'$ and $\nu_\tau'$ separately. The corresponding 
approximate evolution equation is
\be
i\frac{d}{dr}
\left(
\begin{array}{c}
\nu_e \\ \nu_\tau' \\ \bar{\nu}_\mu'
\end{array}
\right)\simeq 
\left(
\begin{array}{ccc}
s_{12}^2 \delta+s_{13}^2\Delta+V_e  & s_{13}\,c_{13}\,\Delta & 
\mu_{e\mu'}B_\perp \\
s_{13}\,c_{13}\,\Delta  & c_{13}^2\,\Delta+V_\mu & \mu_{\mu\tau}B_\perp \\ 
\mu_{e\mu'}B_\perp & \mu_{\mu\tau}B_\perp & c_{12}^2\delta-V_\mu
\end{array}
\right) 
\left( \begin{array}{c}
\nu_e \\ \nu_\tau' \\ \bar{\nu}_\mu'
\end{array}
\right)\,, 
%\equiv H_1 \left( \begin{array}{c}
%\nu_e \\ \nu_\tau' \\ \bar{\nu}_\mu'
%\end{array}
%\right)\,,
\label{evol2}
\ee
where $\Delta\equiv \Delta m_{31}^2/2E$ and $\delta\equiv \Delta 
m_{21}^2/2E$. For small $B_\perp$ the characteristic equation of the  
effective Hamiltonian on the r.h.s. of eq. (\ref{evol2}) can be readily 
solved (note that $\bar{\nu}_\mu'$ decouples in the limit $B_\perp \to 0$), 
and the solutions give the energy levels of matter eigenstates:
%\begin{eqnarray}
\[
 E_{1m,2m}=\frac{1}{2}\left(\Delta+V_e+V_\mu+s_{12}^2\,\delta \pm
\sqrt{(\cos 2\theta_{13}\,\Delta-V_e+V_\mu-s_{12}^2\,\delta)^2
+\sin^2 2\theta_{13}\,\Delta^2 } \right), 
\]
\be
E_{3m}=c_{12}^2\delta-V_\mu\,.
\label{eigen}
%\end{eqnarray}
\ee
We have checked numerically that this approximation is extremely good, and 
for most densities the obtained eigenvalues coincide with those of the 
full (3+3)-state problem to an accuracy better than 1\%.

The MSW-H resonance corresponds to the avoided level crossing of the 
first and second eigenvalues; the resonance density is obtained from the 
condition that the term in brackets under the square root in $E_{1m,2m}$ 
vanishes, which yields
\be
\sqrt{2}\,G_F\frac{1}{m_N}\rho_{res}Y_e=\cos 2\theta_{13}\,\Delta 
-s_{12}^2\,\delta \qquad \qquad\mbox{(MSW-H)}\,.
\label{res4a}
\ee
This equation corrects the 2-flavor resonance condition (\ref{res4}). 
There is one more level 
crossing described by eq. (\ref{evol2}) -- that of the first and third 
eigenvalues (it becomes an avoided level crossing when the magnetic field 
is switched on). The physical interpretation of this level crossing depends 
on where it takes place. If it occurs at $\rho > (\rho_{res})_{\rm MSW-H}$, 
the crossing is on the branch of the first eigenvalue $E_{1m}$ that 
corresponds to $\nu_e$, i.e. it describes the RSFP-E resonance 
$\nu_e\leftrightarrow \bar{\nu}_\mu'$ (see fig. 2a). 
If, however, the crossing occurs at $\rho < (\rho_{res})_{\rm MSW-H}$, 
the resonance point lies on the branch of $E_{1m}$   
corresponding to $\nu_\tau'$. In this case it describes the RSFP-X 
resonance $\nu_\tau'\leftrightarrow \bar{\nu}_\mu'$ (fig. 3a). 
The position of the crossing point depends on the value of $s_{13}$. 
There is the critical value $(s_{13})_c$ for which it occurs at 
$\rho = (\rho_{res})_{\rm MSW-H}$ and which delineates the two 
possibilities: For $s_{13}<(s_{13})_c$ there is the RSFP-X resonance, 
whereas for $s_{13}>(s_{13})_c$ the RSFP-E resonance occurs. 
The critical value $(s_{13})_c$ can be found from eqs. (\ref{eigen}) and 
(\ref{res4a}) and is approximately given in (\ref{cond2}).

The RSFP-E resonance condition (\ref{res6}) is obtained by expanding the 
eigenvalues (\ref{eigen}) in small parameters $1-2Y_e$ and $\delta/\Delta$; 
note, however, that $(s_{13})_c$ cannot be found from the lowest order 
expansion (\ref{res6}) and a more accurate expression has to be used.
With $s_{13}$ varying between $(s_{13})_c\simeq 0.015$ and 0.16, the RSFP-E 
resonance density $\rho_{\rm RSFP-E}$ changes from $\rho_{\rm MSW-H}\simeq 
1.7\times 10^3$ g/cm$^3$ to $1.3\times 10^4$ g/cm$^3$ (for neutrino energy 
$E=15$ MeV).  

The RSFP-X resonance density is in general very close to that of the MSW-H 
resonance: for $s_{13}=(s_{13})_c$ the two resonance densities coincide, 
whereas for $s_{13}\ll (s_{13})_c$ the RSFP-X resonance density is given by  
\be
\sqrt{2}\,G_F\frac{1}{m_N}
\rho_{res}(1-Y_e)\simeq c_{13}^2\Delta - c_{12}^2\delta 
\qquad \qquad\mbox{(RSFP-X)}\,.
\label{res3a}
\ee
This expression corrects the 2-flavor result (\ref{res3}). 

Let us now consider the cases of possible overlap of the resonances. From 
eqs. (\ref{eigen}) and (\ref{res4a}) we find that when the RSFP-E and 
MSW-H (or RSFP-X and MSW-H) resonances are close to each other their 
densities satisfy 
\be
\frac{|\rho_{i}-\rho_{\rm MSW-H}|}{\rho_{\rm MSW-H}} \simeq 
\left|s_{13}^2-(s_{13})_c^2\right|\,\frac{\Delta}{\delta}\frac{1}
{\cos 2\theta_{12}}=
\frac{\left|s_{13}^2-(s_{13})_c^2\right|}{(s_{13})_c}\,, 
\label{ratio}
\ee
where $i=$RSFP-E for $s_{13}>(s_{13})_c$ and $i=$RSFP-X for 
$s_{13}<(s_{13})_c$. 
The condition of no overlap between the MSW-H and RSFP-X resonances can 
be written as \cite{Akh3,APS}
\be
\tan 2\theta_{13}\,(\rho_{res})_{\rm MSW-H} + 
\frac{2\mu_{\mu\tau}B_{\perp r}}{\Delta m_{31}^2/2E}\, 
(\rho_{res})_{\rm RSFP-X}\,< \,(\rho_{res})_{\rm MSW-H}-
(\rho_{res})_{\rm RSFP-X}\,,
\label{nonoverlap1}
\ee
and similar condition can be written for the MSW-H and RSFP-E resonances. 
To find out when the no-overlap conditions are satisfied, let us assume 
that the RSFP-X and RSFP-E resonance widths are small compared to the 
width of the MSW-H resonance, so that the second term on the l.h.s. of 
eq. (\ref{nonoverlap1}) (and the corresponding term in the no-overlap 
condition of the MSW-H and RSFP-E resonances) can be neglected. Then from 
(\ref{nonoverlap1}) and (\ref{ratio}) we find that the conditions 
of no overlap become 
\begin{eqnarray}
\frac{s_{13}}{(s_{13})_c} > \sqrt{2}+1\simeq 2.41\,, \qquad
\mbox{(MSW-H and RSFP-E)}\,; \label{nonoverlap2} \\
\frac{s_{13}}{(s_{13})_c} < \sqrt{2}-1\simeq 0.41\,, \qquad
\mbox{(MSW-H and  RSFP-X)}\,.
\label{nonoverlap3}
\end{eqnarray}
Thus, the RSFP-E and MSW-H (or RSFP-X and MSW-H) resonances overlap 
when $s_{13}$ is within a factor of 2.4 of the critical value $(s_{13})_c
\simeq 0.015$ and do not overlap in the opposite case.

\section{$\nu_e\to \bar{\nu}_e$ conversions of neutronization 
neutrinos }

Consider now the transformations experienced by the neutronization $\nu_e$ 
as they propagate from the supernova's core outwards. 

{\it (a) Normal mass hierarchy}. In this case the $\nu_e\to \bar{\nu}_e$ 
conversion is driven by the \mbox{RSFP-E} resonance or by an interplay of 
the MSW-H and RSFP-X resonances, depending on the value of $s_{13}$.   
Let us first consider the case $s_{13}>(s_{13})_c\simeq 0.015$. The 
neutronization $\nu_e$ first encounter the RSFP-E resonance and, if the 
resonance transition is adiabatic, get converted into $\bar{\nu}_\mu'$ (see 
fig. 2a). As these $\bar{\nu}_\mu'$ propagate towards the region of very 
small matter densities, they transform into the mass eigenstate 
$\bar{\nu}_2$, which contains the $\bar{\nu}_e$ component with the weight 
$s_{12}^2$. Thus, for $s_{13}>(s_{13})_c$ the $\nu_e\to \bar{\nu}_e$ 
conversion probability is 
\be
P(\nu_e\to \bar{\nu}_e)=(1-P_{\rm RSFP-E})\, s_{12}^2\,.
\label{P1}
\ee
In deriving this formula we have assumed that $s_{13}$ is not too close to 
$(s_{13})_c$, so that condition (\ref{nonoverlap2}) is satisfied and the 
\mbox{RSFP-E} and MSW-H resonances do not overlap. We shall consider the case 
of the overlapping resonances below.

In the case $s_{13}<(s_{13})_c\simeq 0.015$, the RSFP-E resonance does not 
exist; the neutronization $\nu_e$ born in the core of the star and 
moving outwards first encounter the MSW-H resonance and then the RSFP-X 
resonance (see fig. 3a). Let us first consider the case when these resonances 
do not overlap; the neutrino conversions in them can then be treated as 
independent. At the MSW-H resonance the neutronization $\nu_e$ get converted 
into $\nu_\tau'$, the efficiency of the transition depending on its degree 
of adiabaticity. The produced $\nu_\tau'$ can then be transformed into 
$\bar{\nu}_\mu'$ in the RSFP-X resonance (see fig. 2a). As these 
$\bar{\nu}_\mu'$ propagate further towards small density regions, they 
transform into the mass eigenstate $\bar{\nu}_2$. Thus, in this case the 
$\nu_e\to \bar{\nu}_e$ conversion probability is 
\be
P(\nu_e\to \bar{\nu}_e)=(1-P_{\rm MSW-H})(1-P_{\rm RSFP-X})\, s_{12}^2\,.
\label{P2}
\ee
The no-overlap condition (\ref{nonoverlap3}) implies $s_{13}^2< [(s_{13})_c 
/2.41]^2\simeq 4\times 10^{-5}$. This means that the MSW-H resonance  
is in fact non-adiabatic or moderately non-adiabatic. 
For the average energy of the neutronization neutrinos $\langle E\rangle\simeq 
15$ MeV, from eqs.  (\ref{res4}), (\ref{gammaMSW2}) and  (\ref{hop}) one finds 
$(1-P_{\rm MSW-H})\lesssim 0.6$, and so the probability $P(\nu_e\to 
\bar{\nu}_e) \lesssim 0.18$.

Let us now consider the case of overlapping resonances. It is instructive 
to calculate the $\nu_e\to \bar{\nu}_e$ conversion probability in 
perturbation theory. Strictly speaking, this approach is only justified 
when $P(\nu_e\to \bar{\nu}_e)\ll 1$; however, it gives the correct 
order-of-magnitude estimate even if this quantity is  not too small. 
Moreover, it allows an exponentiation procedure which is expected to give 
the correct result in a wide range of values of $P(\nu_e\to \bar{\nu}_e)$. 
Another advantage of this method is that it is operative in both 
the overlapping and non-overlapping resonance cases and allows one to 
study the smooth transition from the RSFP-E - mediated $\nu_e\to \bar{\nu}_e$ 
transition to that occurring through the combination of the MSW-H and 
RSFP-X resonances.

We shall make use of the evolution equation (\ref{evol2}). Direct integration 
yields the following expression for the amplitude $\bar{\nu}_\mu'$: 
\be
\bar{\nu}_\mu'(r)=(-i)\int_0^r\!dr' \left[\mu_{e\mu'}\, \nu_e(r') + 
\mu_{\mu\tau}\,\nu_\tau'(r') \right] B_\perp(r')\,
e^{(-i)\int_0^{r'}\!dr_1 (c_{12}^2\delta-V_\mu)}\,.
\label{evol3}
\ee
Here the amplitudes $\nu_e(r)$ and $\nu_\tau'(r)$ have to be found from 
the same system of equations (\ref{evol2}). In perturbation theory one 
considers the $\bar{\nu}_\mu'$ amplitude in the lowest order in magnetic 
field, i.e. neglects the effects of the magnetic field on $\nu_e(r)$ and 
$\nu_\tau'(r)$. The equation for $\bar{\nu}_\mu'(r)$ then decouples  
from the rest of the system, and the amplitudes $\nu_e(r)$ and $\nu_\tau'(r)$ 
can be readily found, e.g., in the adiabatic approximation 
\footnote{This approximation is good in the case of overlapping 
MSW-H and RSFP-X resonances, and also in all cases when the RSFP-E 
resonance transition takes place. The method can be 
easily modified to include possible deviations from adiabaticity.}. 
Eq. (\ref{evol3}) then gives 
\be
\bar{\nu}_\mu'(r)=(-i)\int_0^r\!dr' \left[\mu_{e\mu'}\, s(r') + 
\mu_{\mu\tau} \,c(r') \right] B_\perp(r')\,
e^{-i g(r')}\,.
\label{evol4}
\ee
Here 
\be
g(r)=\int_0^r\!dr_1 \left[ E_{1m}(r_1)-c_{12}^2\delta+V_\mu(r_1)\right]\,,
\label{g}
\ee
and $s(r)$ and $c(r)$ are sine and cosine of the mixing angle   
$\theta_m(r)$ defined through 
\be
\tan 2\theta_m(r)=\frac{\sin 2\theta_{13}\,\Delta}
{\cos 2\theta_{13}\,\Delta-s_{12}^2 \,\delta-
[V_e(r)-V_\mu(r)]}\,.
\label{tan}
\ee
At the MSW-H resonance point, $\theta_m=\pi/4$; at much higher densities 
$\theta_m \simeq \pi/2$, while at densities much lower than the MSW-H 
resonance one, $\theta_m\simeq \theta_{13}\ll 1$.  

The phase factor $e^{-i g(r)}$ in the integrand of (\ref{evol4}) is a 
fast oscillating function for all $r$ except in the vicinity of the 
stationary phase point $r_0$ at which $g'(r_0)=0$; the integral therefore 
receives its main contribution from the neighbourhood of this point and can 
be evaluated in the stationary phase approximation. This gives, up to an 
unimportant phase factor,  
\be
\bar{\nu}_\mu'\simeq 
\sqrt{\frac{2\pi}{|g''(r_0)|}}\,\left[\mu_{e\mu'}\, 
s(r_0) + \mu_{\mu\tau} \,c(r_0) \right] B_\perp(r_0)\,,
\label{stphase}
\ee
where the calculated amplitude $\bar{\nu}_\mu'$ corresponds to large enough 
values of coordinate, so that the neutrinos have already passed through the 
MSW-H and RSFP-E or RSFP-X resonances. The probability of the $\nu_e\to 
\bar{\nu}_e$ conversion is then 
\be
P(\nu_e\to \bar{\nu}_e)=P(\nu_e\to \bar{\nu}_\mu')\,s_{12}^2=
|\bar{\nu}_\mu'|^2\, s_{12}^2\,.  
\label{prob}
\ee

{}From eqs. (\ref{g}) and (\ref{eigen}) one can see that the stationary 
phase condition $g'(r_0)=0$ coincides with the condition of crossing of the 
first and third eigenvalues of the effective Hamiltonian in (\ref{evol2}). 
Thus, the stationary phase point is just the level crossing point that 
defines the positions of the RSFP-E or RSFP-X resonances, depending on 
whether the crossing occurs above or below the MSW-H resonance density. 
Using the stationary phase condition $g'(r_0)=E_{1m}(r_0)-E_{3m}(r_0)=0$, one 
can readily find $g''(r_0)$:
\be
g''(r_0)\simeq 2 V_e\,\frac{s_{13}^2\,\Delta-\cos 2\theta_{12}\,
\delta}{\Delta+(s_{12}^2-2c_{12}^2)\,\delta-2V_e}\,L_\rho^{-1}\,.
\label{gpp1}
\ee  
Here we have taken into account that $Y_e\simeq 1/2$. 

Consider first the case $s_{13}-(s_{13})_c \gg 
(s_{13})_c$. This gives $\rho(r_0)\gg \rho_{\rm MSW-H}$, i.e. $2V_e(r_0) \gg 
\Delta$. From eq. (\ref{gpp1}) we then find 
\be
g''(r_0)\simeq -L_\rho^{-1}\,(s_{13}^2\,\Delta-\cos 2\theta_{12}\,\delta)\,.
\label{gpp2}
\ee
Substituting this into eq. (\ref{stphase}) one obtains 
\be
P(\nu_e\to\bar{\nu}_\mu')=
|\bar{\nu}_\mu'|^2\simeq 
\frac{\pi}{2}\,\frac{4 \left[\mu_{e\mu'}\, B_\perp(r_0)\right]^2}
{s_{13}^2\,\Delta-\cos 2\theta_{12}\,\delta}\,L_\rho\,,
\label{P3}
\ee
where we have taken into account that for $\rho(r_0)\gg \rho_{\rm MSW-H}$ 
one has $s(r_0)\simeq 1$, $c(r_0)\simeq 0$. Notice that this expression 
coincides with $(\pi/2)\gamma_{\rm RSFP-E}$ where $\gamma_{\rm RSFP-E}$ 
was defined in eq. (\ref{gammaRSFPE}). Thus, the $\nu_e\to \bar{\nu}_\mu'$ 
conversion in this case is driven by the RSFP-E resonance. It is interesting 
to note that the perturbation-theoretic expression (\ref{P3}) is just the 
first term in the expansion of the transition probability $P(\nu_e\to 
\bar{\nu}_\mu')=1-\exp{[-(\pi/2)\gamma_{\rm RSFP-E}]}$ in the small 
$\gamma_{\rm RSFP-E}$ limit. Thus, eq. (\ref{P3}) is in accord with eqs. 
(\ref{P1}) and (\ref{prob}). 

Let us now consider the case $s_{13}\ll (s_{13})_c$. In this case $\rho(r_0)$ 
coincides with the small-$s_{13}$ limit of $\rho_{\rm RSFP-X}$ given in eq. 
(\ref{res3a}). Using this condition we find from (\ref{gpp1})
\be
g''(r_0)\simeq L_\rho^{-1}\,(c_{13}^2\,\Delta-c_{12}^2\,\delta)\,.
\label{gpp3}
\ee
Substituting this into eq. (\ref{stphase}) yields  
\be
P(\nu_e\to\bar{\nu}_\mu')=|\bar{\nu}_\mu'|^2\simeq 
\frac{\pi}{2}\,\frac{4 \left[\mu_{\mu\tau}\, B_\perp(r_0)\right]^2}
{c_{13}^2\,\Delta-c_{12}^2\,\delta}\,L_\rho\,,
\label{P4}
\ee
where we have taken into account that for densities below $\rho_{\rm MSW-H}$ 
and outside the RSFP-H resonance region one has $s(r_0)\simeq 0$, $c(r_0) 
\simeq 1$. We note that this expression coincides with 
$(\pi/2)\gamma_{\rm RSFP-X}$. Thus, the $\nu_e\to \bar{\nu}_\mu'$ conversion 
in this case is driven by the combination of the MSW-H and RSFP-X 
resonances and goes through the chain of transitions $\nu_e\to\nu_\tau'\to 
\bar{\nu}_\mu'$. Eq. (\ref{P4}) gives the first term in the expansion of the 
transition probability 
\be
P(\nu_e\to \bar{\nu}_\mu')= (1-e^{-\frac{\pi}{2}\gamma_{\rm 
MSW-H}})(1-e^{-\frac{\pi}{2}\gamma_{\rm RSFP-X}})
\label{PP}
\ee 
in small $\gamma_{\rm RSFP-X}$ (note that in our case the first factor 
in (\ref{PP}) is equal to unity since we assume the MSW-H transition to 
be adiabatic). Thus, eq. (\ref{P4}) agrees with eqs. (\ref{P2}) and 
(\ref{prob}).

We have considered the cases when the stationary point is situated above 
or below the MSW-H resonance point and sufficiently far from it. We found 
that in those cases the $\nu_e\to \bar{\nu}_\mu'$ transition is driven 
either by the RSFP-E resonance or by the sequence of the \mbox{MSW-H} and 
RSFP-X resonances, as the resonance regions do not overlap. When $s_{13}
\simeq (s_{13})_c$, the stationary phase point is close to the MSW-H 
resonance one, which leads to the overlap of the resonances. In that 
case the transition mechanism is a subtle interplay of both the mechanisms 
discussed above. In particular, when $s_{13}=(s_{13})_c$ one has 
$s(r_0)=c(r_0)=1/\sqrt{2}$, and the magnetic moments $\mu_{e\mu'}$ and 
$\mu_{\mu\tau}$ enter into the amplitude $\bar{\nu}_\mu'$ in eq. 
(\ref{stphase}) with equal weights. This corresponds to the maximal 
interference of the two mechanisms discussed above; the interference can 
be either constructive or destructive, depending on the relative sign of
$\mu_{e\mu'}$ and $\mu_{\mu\tau}$. When $s_{13}\simeq (s_{13})_c$ but not 
exactly equal to this critical value, the direct $\nu_e\to \bar{\nu}_\mu'$ 
conversion due to the RSFP-E resonance and the $\nu_e\to\nu_\tau'\to 
\bar{\nu}_\mu'$ transitions due to the combined action of the MSW-H and 
RSFP-X resonances give comparable contributions to $P(\nu_e\to 
\bar{\nu}_\mu')$. The relative weight of these contributions depends on 
the value of the mixing angle $\theta_m$ at the level crossing point $r_0$.

We have found that in the cases of non-overlapping resonances the 
perturbation-theoretic expressions for the probability $P(\nu_e\to 
\bar{\nu}_\mu')$ are just the first-order terms in the expansions of the 
expressions of the type $(1-\exp[-(\pi/2)\gamma_i])$. It is therefore natural 
to assume that this is also true in general, and the probability $P(\nu_e\to 
\bar{\nu}_\mu')$ can be found from (\ref{stphase}) by the exponentiation 
procedure. According to this procedure, the probability 
$P(\nu_e\to \bar{\nu}_\mu')$ is obtained as 
\be
P(\nu_e\to \bar{\nu}_\mu')=1-e^{-\frac{\pi}{2}\gamma}\,,
\label{P5}
\ee
where 
\be
\gamma=\frac{4}{|g''(r_0)|}\, \left[\mu_{e\mu'}\, s(r_0)+\mu_{\mu\tau} 
\,c(r_0) \right]^2 B_\perp(r_0)^2\, 
\label{gamma}
\ee
with $g''(r_0)$ defined in eq. (\ref{gpp1}). Eqs. (\ref{P5}), (\ref{gamma}) 
are expected to be valid for all values of $\gamma$, not necessarily 
$\gamma \ll 1$, provided that the MSW-H resonance conversion is 
adiabatic. The probability of the $\nu_e\to \bar{\nu}_e$ conversion we 
are interested in is then found from eq. (\ref{prob}).

As can be seen from eqs. (\ref{P1}), (\ref{P2}) and (\ref{prob}), in the 
case of the normal neutrino mass hierarchy the maximum possible value of 
$P(\nu_e\to \bar{\nu}_e)$ is $s_{12}^2\simeq 0.3$.

{\it (b) Inverted mass hierarchy.} In this case the neutronization $\nu_e$ 
first encounter the \mbox{RSFP-H} resonance and get converted into 
$\bar{\nu}_\tau'$. The destiny of the latter depends on whether $s_{13}$ 
is above or below the critical value $(s_{13})_c$ (see figs. 2b and 3b).

Let us first consider the case $s_{13}>(s_{13})_c$. The $\bar{\nu}_\tau'$ 
that emerge from the RSFP-H resonance next reach the MSW-H resonance and 
get converted into $\bar{\nu}_e$, which then encounter the RSFP-E resonance 
where they can be transformed into $\nu_\mu'$ (fig. 2b). If this latter 
conversion occurs, the neutronization $\nu_e$ end up as neutrinos 
and not antineutrinos (in general, for a $\nu_a \to \bar{\nu}_b$ conversion, 
neutrinos should experience an odd number of the RSFP transformations). 
Therefore the $\nu_e\to\bar{\nu}_e$ transition we are interested in takes 
place only if the RSFP-E resonance is non-adiabatic. Then $\bar{\nu}_e$  
emerging from the MSW-H resonance pass through the RSFP-E resonance 
unaffected.  As they propagate towards very small densities, 
they transform into $\bar{\nu}_1$, which have the $\bar{\nu}_e$ component 
with the weight $c_{12}^2$. Thus, the $\nu_e\to\bar{\nu}_e$ conversion 
occurs through the sequence of transitions $\nu_e\to\bar{\nu}_\tau'\to
\bar{\nu}_e\longrightarrow \bar{\nu}_2$, and its probability is
\be
P(\nu_e\to \bar{\nu}_e)=(1-P_{\rm RSFP-H})(1-P_{\rm MSW-H})\,
P_{\rm RSFP-E}\,c_{12}^2\,.
\label{P6}
\ee

In the case $s_{13}<(s_{13})_c$, the $\bar{\nu}_\tau'$ emerging from the 
RSFP-H resonance next reach the RSFP-X resonance and can be converted there 
into $\nu_\mu'$ (fig. 3b). For the $\nu_e\to\bar{\nu}_e$ transition to 
occur, this resonance has to be non-adiabatic. Then $\bar{\nu}_\tau'$ 
pass through the RSFP-X resonance unscathed and propagate towards the 
\mbox{MSW-H} resonance where, as in the case $s_{13}>(s_{13})_c$, they can 
be converted into $\bar{\nu}_e$. These 
$\bar{\nu}_e$ subsequently transform into $\bar{\nu}_1$ in the small 
density regions. The sequence of transitions in this case is therefore the 
same as in the case $s_{13}>(s_{13})_c$. The probability of the $\nu_e\to
\bar{\nu}_e$ conversion is 
\be
P(\nu_e\to \bar{\nu}_e)=(1-P_{\rm RSFP-H})(1-P_{\rm MSW-H})\,
P_{\rm RSFP-X}\,c_{12}^2\,.
\label{P7}
\ee
Note that in the case $s_{13}<(s_{13})_c$, the MSW-H resonance transition 
is adiabatic only for $s_{13}$ relatively close to $(s_{13})_c$.

As can be seen from eqs. (\ref{P6}) and (\ref{P7}), in the case of the 
inverted neutrino mass hierarchy the maximum possible value of $P(\nu_e\to 
\bar{\nu}_e)$ is $c_{12}^2\simeq 0.7$.

\section{Discussion}
We have considered neutrino flavor and spin-flavor transitions in supernovae 
in the full \mbox{3-flavor} framework and found that, in addition to the 
known MSW and RSFP resonances that can be obtained assuming the transitions 
to be approximately 2-flavor ones, there are new RSFP resonances which 
are pure 3-flavor effects and cannot be found in 2-flavor approximations. 
We have studied these new resonances and their interplay with the other 
nearby resonances in some detail, including the case of overlapping 
resonances. We have explored the role of these resonances in the 
transformation of neutronization $\nu_e$ into their antiparticles. It was 
found that such transformations depend crucially on the value of the neutrino 
mixing parameter $s_{13}$ and are in general possible for both normal and inverted 
neutrino mass hierarchies. We obtained the relevant transition probabilities 
in each case. 

Let us discuss now the conditions on neutrino magnetic moments and SN 
magnetic fields that have to be satisfied in order for the $\nu_e\to
\bar{\nu}_e$ transitions of the neutronization neutrinos to be efficient. 
In the case of the normal neutrino mass hierarchy the $\nu_e\to\bar{\nu}_e$ 
transition probabilities are given by eqs. (\ref{P1}) and (\ref{P2}) for 
$s_{13}>(s_{13})_c$ and $s_{13}<(s_{13})_c$ respectively. For $s_{13}>
(s_{13})_c$ the efficiency of the $\nu_e\to\bar{\nu}_e$ conversion is 
determined by the RSFP-E adiabaticity parameter, eq. (\ref{gammaRSFPE}). 
Assuming the power-law magnetic field profile (\ref{B}), we find from eqs. 
(\ref{condit1}) and (\ref{condit2}) that in the case $B_0=10^{14}$ G and $k=2$ 
the transition is adiabatic ($\gamma_{\rm RSFP-E}>1$) if $\mu_{e\mu'}> 
5\times 10^{-13}\mu_B$, while for the exponent $k=3$ this would require 
$\mu_{e\mu'}> 10^{-9}\mu_B$, a value already experimentally excluded. Note 
that magnetic fields as strong as $10^{16}$ G have been considered possible 
in supernovae \cite{supB}; if this is the case, for $k=3$ the transition 
magnetic moments $\mu_{e\mu'}\sim 10^{-11}\mu_B$ would cause strong $\nu_e\to
\bar{\nu}_e$ conversions, while for $k=2$ magnetic moments as small as 
$5\times 10^{-15}\mu_B$ would do. 

For $s_{13}<(s_{13})_c$ the $\nu_e\to\bar{\nu}_e$ conversion is driven by 
a combination of the MSW-H and RSFP-X resonance transitions. For values 
of $s_{13}$ only slightly below the critical value, the transition 
efficiency is mainly determined by the RSFP-X adiabaticity parameter.  
{}From eqs. (\ref{adiab2}) and (\ref{adiab5}) we find that in the case 
$B_0=10^{14}$ G and $k=2$ the RSFP-X transition is adiabatic if 
$\mu_{\mu\tau}> 10^{-11}\mu_B$, while for the exponent $k=3$ this would 
require $\mu_{\mu\tau}>4\times 10^{-8}\mu_B$. For different values of $B_0$ 
these limits would have to be rescaled accordingly. 
%%%%%%%%%%%%%%%%%%%%%%%%%%%%%%%%%%%%%%%%%%%%%%%%%%%%%%%%%%
%Note that the laboratory 
%upper bounds on the transition magnetic moments $\mu_{\mu\tau}$ are much 
%less stringent than those on $\mu_{e\mu}$ and $\mu_{e\tau}$ \cite{PDG}; 
%however, the astrophysical limits obtained from star cooling rates 
%\cite{cool} apply equally to all transition magnetic moments. 
%%%%%%%%%%%%%%%%%%%%%%%%%%%%%%%%%%%%%%%%%%%%%%%%%%%%%%%%%%

In the case of the inverted neutrino mass hierarchy the $\nu_e\to\bar{\nu}_e$ 
transition probabilities are given by eqs. (\ref{P6}) and (\ref{P7}) for 
$s_{13}>(s_{13})_c$ and $s_{13}<(s_{13})_c$ respectively. If the MSW-H 
transition is adiabatic, the $\nu_e\to\bar{\nu}_e$ conversion probability is 
determined by the RSFP-H adiabaticity parameter $\gamma_{\rm RSFP-H}$.  
{}From eqs. (\ref{adiab1}) and (\ref{adiab3}) we find that for $B_0=10^{14}$ 
G and $k=2$ the transition is adiabatic if $\mu_{e\tau'}> 
5\times 10^{-13}\mu_B$, while for the exponent $k=3$ the adiabaticity of 
the transition would require $\mu_{e\tau'}> 2\times 10^{-10}\mu_B$. These 
conditions are comparable with the conditions we obtained for $\mu_{e\mu'}$ 
in the case of the normal neutrino mass hierarchy and $s_{13}>(s_{13})_c$. 
From eqs. (\ref{P6}) and (\ref{P7}) it follows that in the inverted hierarchy 
case the $\nu_e\to\bar{\nu}_e$ transitions can only be efficient if the 
RSFP-X transition [for $s_{13}< (s_{13})_c)$] or RSFP-E transition [for 
$s_{13}> (s_{13})_c)]$ is non-adiabatic. It is easy to see that these 
conditions can be satisfied without any conflict with the requirement of 
the adiabaticity of the RSFP-H condition. 

It might also be useful to give the conditions for the efficient $\nu_e\to
\bar{\nu}_e$ transitions of the neutronization neutrinos directly in terms
of the SN magnetic field strength at the resonance, i.e. independently of the
supernova magnetic field model. In the case of the normal neutrino mass  
hierarchy and $s_{13}>(s_{13})_c$, the requirement that the RSFP-E   
adiabaticity parameter exceeds unity yields $\mu_{e\mu'} B_{\perp r}>10^{-13}
\mu_B\times 1.5\cdot 10^8$ G. For the normal hierarchy and $s_{13}$ slightly
below the critical value $(s_{13})_c$, the condition $\gamma_{\rm RSFP-X}>1$
leads to $\mu_{\mu\tau} B_{\perp r}> 10^{-13}\mu_B\times 1.5\cdot 10^9$~G. In
the case of the inverted neutrino mass hierarchy the $\nu_e\to\bar{\nu}_e$
transitions can only be efficient when $\gamma_{\rm RSFP-H}>1$, which   
yields $\mu_{e\tau'} B_{\perp r}>10^{-13}\mu_B\times 4\cdot 10^{9}$ G.

Thus, we have seen that in both cases of the normal and inverted neutrino 
mass hierarchies sizeable $\nu_e\to\bar{\nu}_e$ transitions of the SN 
neutronization neutrinos are in general possible. In the normal hierarchy 
case, the maximal possible transition probability is equal to $s_{12}^2
\simeq 0.3$, whereas for the inverted hierarchy it is $c_{12}^2\simeq 0.7$. 
Thus, in the inverted hierarchy case the $\nu_e\to\bar{\nu}_e$ conversion 
probability can be higher. In both cases the $\nu_e\to\bar{\nu}_e$ conversion 
occurs only when $s_{13}$ is not too small: For the RSFP-E -- driven 
transitions, one needs $s_{13}> (s_{13})_c\simeq 0.015$ for the RSFP-E 
resonance to exist; for the transitions occurring through the combinations
of the MSW-H and RSFP-X resonances or RSFP-H and MSW-H resonances, one 
needs $s_{13}\gtrsim 10^{-2}$ for the MSW-H resonance to be adiabatic. 

Conversion of the SN neutronization $\nu_e$ into $\bar{\nu}_e$ would 
have a very clear and distinct experimental signature, especially in water 
Cherenkov detectors. We will consider the Super-Kamiokande detector (fiducial 
volume 32 kt) as an example. For a galactic supernova at 10 kpc from the 
Earth, one expects in this detector about 12 events from the detection of 
the neutronization $\nu_e$ through the $\nu_e e^- \to\nu_e e^-$ scattering 
reaction, assuming no flavor or spin-flavor conversions (see fig. 22 in 
\cite{ThBuPi}). These events should occur in a very short time interval of 
$\sim$ (10 -- 20) ms and should precede a longer signal of neutrinos and 
antineutrinos of all flavors. If the $\nu_e\to\bar{\nu}_e$ conversion occurs, 
the neutronization burst observed by terrestrial detectors should contain a 
significant fraction of electron antineutrinos, up to 30\% in the normal 
hierarchy case and up to 70\% in the case of the inverted mass hierarchy. 
The main detection mechanism of $\bar{\nu}_e$ is through the $\bar{\nu}_e p
\to n e^+$ reaction, which has 
a much larger cross section than that of $\nu_e e^-$ scattering: at the 
average energy of the neutronization neutrinos $\langle E\rangle \simeq 15$ 
MeV one has $\sigma(\bar{\nu}_e p\to n e^+)/\sigma(\nu_e e^- \to\nu_e e^-)
\simeq 150$. Therefore one can expect a very strong signal of $\bar{\nu}_e$ 
-- up to 500 (1200) events in the case of the normal (inverted) mass 
hierarchy -- in a very short time interval of $\sim$ (10 -- 20) ms. Note 
that $\bar{\nu}_e$ can be cleanly distinguished experimentally from all other 
neutrino and antineutrino species \cite{VB}.

Supernova neutronization $\bar{\nu}_e$ can also be observed  
in the SNO detector, which contains about 1.4 kt of light water in addition 
to 1 kt of heavy water. The $\bar{\nu}_e p\to n e^+$ capture reaction in 
light water can result in up to 20 (50) events in the case of the normal
(inverted) neutrino mass hierarchy. Detection of the neutronization 
$\bar{\nu}_e$ through the charged-current reaction $\bar{\nu}_e d\to n n e^+$  
in heavy water is of lesser interest since the cross section of this 
process is about a factor of three smaller than that of the $\bar{\nu}_e p$ 
capture. Still, the $\bar{\nu}_e d\to n n e^+$ events at SNO can be a 
useful complement to the $\bar{\nu}_e p\to n e^+$ ones.

In the beginning of this section we estimated the neutrino transition 
magnetic moments and SN magnetic fields which are necessary for appreciable 
$\nu_e\to\bar{\nu}_e$ conversions to take place. These estimates were 
obtained assuming that the relevant RSFP adiabaticity parameters satisfy 
$\gamma_{\rm RSFP}=1$, which corresponds to the transition probability 
of about 80\%. Therefore for these values of the neutrino magnetic moments 
and SN magnetic fields the above estimates of the expected numbers of 
$\bar{\nu}_e$ events have to be reduced by the factor 0.8. On the other 
hand, if, say, 30 $\bar{\nu}_e$ events in Super-Kamiokande can be considered 
as a clear and unambiguous signal, the requisite values of the transition 
magnetic moments will be reduced by a factor of 5 (8) for the normal 
(inverted) neutrino mass hierarchy as compared to our previous estimates. 
Future very large SN neutrino detectors would have an even better sensitivity 
to neutrino magnetic moments and SN magnetic fields. 

Let us discuss now the model dependence of our results. The properties 
of the SN neutronization pulse calculated by different groups are 
in a reasonably good agreement, as can be seen, e.g., from the comparison 
of refs. \cite{ThBuPi} and \cite{LRJM}. Thus, they can be considered as 
relatively well known. The same applies to the supernova density profiles 
$\rho(r)$, for which different groups converge at similar $1/r^3$ dependences  
\cite{BBB,JH,WW}. The quantity $1-2Y_e$ in the region where most of the 
resonances occur is less well established, though; it depends strongly on 
the assumed metallicity of the pre-supernova model. The dependence of the 
RSFP of SN neutrinos on the SN models, including the density and $1-2Y_e$ 
profiles, was studied in \cite{SNmodeldep}; in most cases relatively mild 
model dependence was found.

The main uncertainty in our results is related to the fact that the strengths 
and profiles of the SN magnetic fields are essentially unknown. For this 
reason, in addition to studying the two popular power-law profiles, we 
expressed our results directly in terms of the magnetic field strength at the 
resonance, thus reducing their model dependence.

Is RSFP the sole mechanism that can convert the SN neutronization $\nu_e$ 
into $\bar{\nu}_e$? In principle, SN neutrinos could also experience 
$\nu_e\to\bar{\nu}_e$ transitions due to a combination of the MSW conversions 
and neutrino decay into a lighter (anti)neutrino and Majoron 
\cite{CMP,SVa,BV,BS,KL,KTV,LOW,Farzan}. 
For hierarchical neutrino masses such a decay  would lead to a significant 
degradation of the neutrino energy, and so the $\nu_e\to \bar{\nu}_e$ 
conversion due to the neutrino transition magnetic moments would be clearly 
distinguishable experimentally from that caused by the neutrino decay.  
The situation is more complicated if the decaying and daughter neutrinos 
are quasi-degenerate in mass. In this case there will be essentially no 
neutrino energy degradation, as was recently emphasized in \cite{BeBe}. 
However, in the case of decaying SN neutrinos the decay of thermally 
produced neutrinos would result in composite spectra of the detected  
neutrinos which would be different from those expected in the case of the pure 
MSW effect. The decaying neutrino scenario can also be independently tested 
through the decay of high-energy astrophysical neutrinos \cite{BBHPW1,BBHPW2}.

To conclude, the $\nu_e\rightarrow \bar{\nu}_e$ conversion of supernova 
neutrinos due to the combined action of neutrino flavor mixing and transition 
magnetic moments can lead to an observable signal of the neutronization 
neutrinos in the $\bar{\nu}_e$ channel. Such an effect would have a clear 
experimental signature and its observation would be a smoking gun evidence 
for the neutrino transition magnetic moments. It would also signify a 
relatively large leptonic mixing parameter $|U_{e3}|=s_{13}> 10^{-2}$.

\vspace*{2mm}
{\em Note added.} While this paper was in preparation, the papers 
\cite{AhMi,AS} appeared in which the $\nu_e\to \bar{\nu}_e$ conversions 
of the supernova neutronization neutrinos were also considered. In the 
case of the inverted mass hierarchy, our conclusions essentially coincide 
with those in \cite{AhMi,AS}; however, the authors of those papers did not 
take into account the RSFP-E and RSFP-X resonances and therefore missed the 
possibility of a strong $\nu_e\to \bar{\nu}_e$ conversion in the case of the 
normal neutrino mass hierarchy, which was studied in the present paper.

{\em Acknowledgments.} The authors are grateful to Alexei Smirnov 
for useful discussions. TF acknowledges the hospitality of the Abdus 
Salam International Centre of Theoretical Physics, Trieste, where this
work was initiated.

\end{document}